# Quantum Key Distribution in a Multi-User Network at Gigahertz Clock rates

Veronica Fernandez*[a], Karen J. Gordon[a], Robert J. Collins[a], Paul D. Townsend[b], Sergio D. Cova[c], Ivan Rech[c], Gerald S. Buller[a]

[a] School of Engineering and Physical Sciences, Heriot-Watt University, Edinburgh, UK, EH14 4AS
[b] Photonics Systems Group, Physics Department, University College Cork, Cork, Ireland
[c] Dipartimento Elettronica e Informazione, Politecnico di Milano, 20133, Milano, Italia

## ABSTRACT

In recent years quantum information research has lead to the discovery of a number of remarkable new paradigms for information processing and communication. These developments include quantum cryptography schemes that offer unconditionally secure information transport guaranteed by quantum-mechanical laws. Such potentially disruptive security technologies could be of high strategic and economic value in the future. Two major issues confronting researchers in this field are the transmission range (typically <100km) and the key exchange rate, which can be as low as a few bits per second at long optical fiber distances. This paper describes further research of an approach to significantly enhance the key exchange rate in an optical fiber system at distances in the range of 1-20km. We will present results on a number of application scenarios, including point-to-point links and multi-user networks.

Quantum key distribution systems have been developed, which use standard telecommunications optical fiber, and which are capable of operating at clock rates of up to 2GHz. They implement a polarization-encoded version of the B92 protocol and employ vertical-cavity surface-emitting lasers with emission wavelengths of 850 nm as weak coherent light sources, as well as silicon single-photon avalanche diodes as the single photon detectors. The point-to-point quantum key distribution system exhibited a quantum bit error rate of 1.4%, and an estimated net bit rate greater than 100,000 bits$^{-1}$ for a 4.2 km transmission range.

**Keywords:** Quantum key distribution, quantum cryptography, optical fiber communications, data security.

## 1. INTRODUCTION

In this era of information technology the need for secure communications has become increasingly important. Taking advantage of the Heisenberg's Uncertainty Principle, which states that in general we cannot obtain full information on a quantum system as we perform a measurement on it, Quantum Key Distribution (QKD) provides a unique secure way to share an encryption key between two users. The "one time pad" approach [1] can utilize this key to provide absolute security of encryption, provided that the key is random, is as long as the message and is used only once [2]. QKD exploits the fundamental laws of quantum mechanics to achieve security in key distribution: for example, the fact that the non-orthogonal polarization states of individual photons cannot be simultaneously measured with arbitrarily high accuracy. Hence, if the random key data is encoded and transmitted using such states, secure key sharing can be achieved because an eavesdropper can only ever obtain partial information on the key and will inevitably generate a detectable disturbance on the quantum communication channel. Bennett and Brassard's original QKD protocol (BB84) [3] employs two incompatible pairs of conjugate quantum observables, for example circular and linear-polarization states, to encode the data. In Bennett's simpler B92 protocol two non-orthogonal states are utilized, for example two non-orthogonal linear-polarization states [4]. To date, many groups have demonstrated the possibility of free-space [5,6] and optical fiber-based QKD systems [7,8]. However the key exchange in those systems remains, in general, below 1kbits$^{-1}$, in particular for λ ~ 1.55 µm QKD systems, when InGaAs/InP is used as the single-photon detectors. This is due to the frequency restriction imposed by the deleterious effects of afterpulsing phenomenon, which is present in these detectors even at low count rates. A different approach is possible to increase the bit rate, which consists of using Si Single-Photon Avalanche Diode (SPAD) detectors in conjunction with standard telecommunication fiber [9]. In this paper, we present a modification of the QKD system described in [9] to include an electronically enhanced commercially-available silicon single-photon counting module (SPCM), allowing faster clock rates to be employed. We show that the use of the enhanced module in the QKD system enables operation at clock rates of up to 2GHz [10]. We also describe an investigation into the use of a 1×32 network [11] – i.e. one transmitter of the initial photon stream and 32 independent potential receivers. Both systems were characterized in terms of quantum bit error rate (QBER) and estimated net bit rate (NBR) of the final shared key, as discussed previously in [9].

## 2. DESCRIPTION OF THE SYSTEM

The characterized QKD system is shown in Fig.1. It is entirely fiber-based, and an interchangeable length of standard telecommunications fiber separates Alice (the transmitter) and Bob (the receiver).

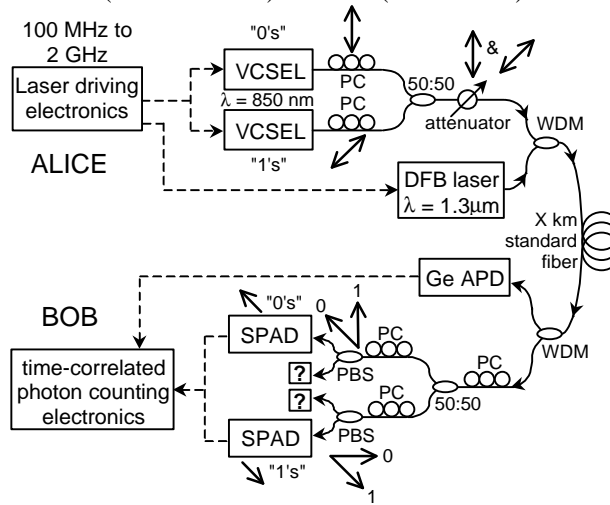

Fig. 1. Schematic diagram of the quantum key distribution experiment. PBS: Polarization splitter. PC: Polarization controller. WDM: Wavelength division multiplexor. APD: Avalanche photodiode. VCSEL: Vertical-cavity surface-emitting laser. SPAD: Single-photon avalanche diode. DFB: Distributed feedback laser. The arrows represent polarization states of individual photons propagating in the optical fiber.

Two 850 nm wavelength vertical-cavity surface-emitting lasers (VCSELs) were used to implement the B92 protocol. The VCSELs were driven using two independent laser driver circuit boards, which enabled single mode operation when driven under the correct conditions (< 7mA drive current). The laser driver boards were driven using a non-return to zero (NRZ) differential output from a preprogrammed pulse pattern generator (PPG). In order to reduce the probability of a multiple photon event in any pulse period, the combined VCSEL pulses were attenuated to an average of ~0.1 photons per pulse. This ensured that the probability of two photons occurring in any pulse period was less than 0.5%, thus reducing the likelihood of a successful photon number splitting eavesdropping attack [12].

The polarized light from each VCSEL was launched into 5.5μm core diameter optical fiber, which was single mode at a wavelength of 850 nm. Both Alice and Bob utilized 850-nm single-mode fiber, but were separated by an interchangeable length of standard telecommunications fiber. The latter supports more than one mode at an operating wavelength of 850 nm. However, when a length of standard telecommunications fiber is fusion spliced to a fiber of smaller core diameter the fundamental mode is excited and the second order modes are greatly suppressed. Thus, only one stable spatial mode propagates in the fiber at all times [13].

Two polarization controllers were used to set the polarization states in the two channels to linear with a relative polarization angle of 45°. The non-orthogonal polarization states were combined by a polarization independent 50:50 coupler, and attenuated to a mean number of 0.1 photons per laser pulse (μ) using a programmable attenuator.

The VCSELs were temperature tuned to have identical emission wavelengths of 850nm in order to avoid spectral interrogation by an eavesdropper. The system was optically synchronized using intense pulses (~$1.5 \times 10^8$ photons per pulse) generated by a 1.3μm wavelength distributed feedback (DFB) laser, which were wavelength multiplexed into the fusion spliced transmission fiber connecting Alice and Bob. The two wavelengths were demultiplexed at Bob, and the 1.3μm wavelength pulses were detected and converted into an electrical signal by a linear gain germanium avalanche photodiode (APD), biased ~1 V below avalanche breakdown (in the linear multiplication regime). The output of the Ge APD was then directed to the synchronization input of the photon-counting acquisition card.

A bandpass filter (Δλ = 30nm centered at 850nm) was inserted into the 850nm quantum channel in order to block any remaining 1.3μm wavelength light not removed by the demultiplexer. The dual polarization states arriving at Bob after propagating through $X$ km of standard telecommunications fiber, were manipulated to enable 25% of the transmitted data to be measured unambiguously. Two fiber based polarizing beam splitters (PBSs) and two silicon SPADs were used for this purpose.

We will denote the fiber channel from which the unambiguous "0s" are measured as channel 0, and correspondingly the channel from which the unambiguous "1s" are measured as channel 1. 1. Note that there are two other fiber ports from the polarizing beam splitters indicated in fig. 1 by a "?". These ports indicate the two fiber channels that contain the remaining 75% of the ambiguous data. The photons in these ports could be measured using another two SPADs, and the collected data could be analyzed statistically to detect the presence of an eavesdropper [14].

The "unambiguous" photons from channel 1 and channel 0 were then detected using two commercially-available SPCM-AQR single-photon detection modules obtained from Perkin Elmer (PKI), one of which was modified for improved performance as described below. The output from the SPCM modules was directed to the photon-counting acquisition card, which can simultaneously acquire data from both channels. Data was then collected using the detected photons and synchronization pulse in order to characterize the system performance in terms of quantum bit error rate (QBER).

## 3. ENHANCED DETECTOR

The technique described in [15,16] was exploited to improve the photon timing performance of a standard PKI SPCM-AQR photon detector module. An additional circuit card [16] was inserted into the module without modifying the original circuit card, which still quenched the avalanche and could still be used for pulse counting. This modification was reversible and the additional circuit card could be removed with no detrimental effects to the original circuit card. A pulse pick-off linear network was connected to the SPAD terminal, which was biased at a high-voltage (about 400V). The network was specifically designed to extract a short pulse signal with fast rise, practically coincident with the rise of the avalanche current. A fast discriminator with very low sensing threshold was then employed for sensing the onset of this pulse. Therefore, the avalanche current can be sensed at an initial stage of its build-up, when still confined in a small area of the detector. Thus, the time information obtained was not affected by the statistical fluctuations that characterize the propagation of the current over the full area of the detector [17]. Hence, the jitter in the measured arrival time of the photon was minimized.

Prior to modification the original module had a full-width at half-maximum (FWHM) jitter of ~570ps at low counting rates. With the modification the module had an improved FWHM jitter of ~370ps. The additional circuit card has the added advantage that its performance is more stable at high pulse counting rates (typically above 0.5Mcounts$^{-1}$). At such rates it is important that the recovery to the baseline level after an avalanche pulse is fast and accurate. If this recovery has a slower tail, even one with a small amplitude, then the superposition of such tails will cause both fluctuations and a systematic mean shift of the baseline level, which causes random fluctuations and shift of the triggering level along the pulse risetime. The corresponding effect on the measured photon arrival time causes a degradation of the FWHM value and a systematic shift of the centroid and of the peak of the photon timing distribution. These effects, which are significant with the original PKI circuit, are strongly reduced by the additional circuit card, as illustrated in figures 2 and 3. For example, at an incident count rate of 2Mcounts$^{-1}$ the modified device exhibits a jitter of ~450ps (FWHM), compared with ~950ps jitter prior to modification. Temporal broadening of the single-photon detector response has been shown to limit the performance of the QKD system [9] since at clock frequencies between 1 and 2GHz and short fiber lengths the detected count rate can be between 0.5 and 1.5Mcounts$^{-1}$. The reduction in the peak shift does not directly improve the QBER, it does however allow the data collection window to stay fixed with respect to the synchronization pulse [9]. In a high bit rate transmission system this is of fundamental importance, since it means that each bit remains within the proper time slot allotted to it.

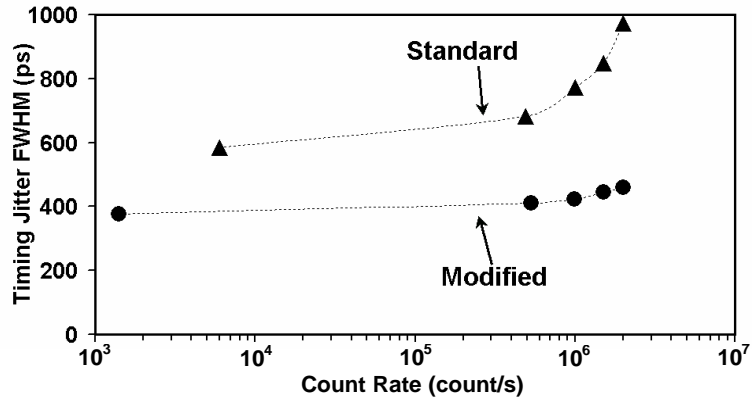

Fig. 2. Timing jitter full width at half maximum of the standard SPCM SPAD and the SPCM SPAD with modified output circuitry.

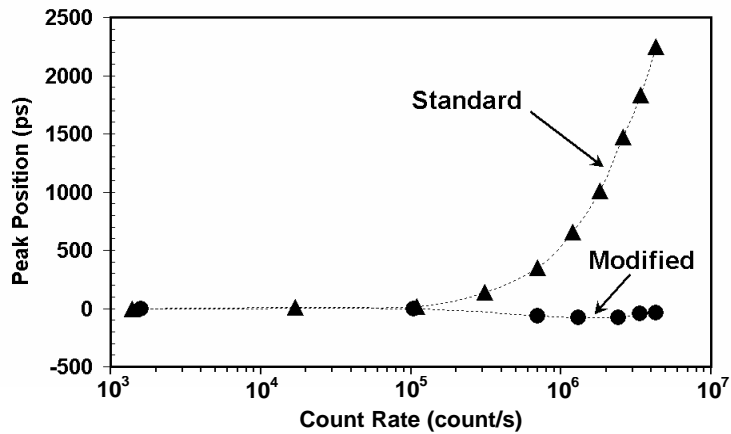

Fig. 3. Shift of the peak position of the standard SPCM SPAD and the SPCM SPAD with modified output circuitry.

### 3.1. QKD SYSTEM EXPERIMENTAL RESULTS WITH ENHANCED DETECTOR

In this section we show significant improvements in experimental data in terms of QBER by comparing data taken using the standard output of a Perkin Elmer SPCM-AQR photon detector and the output from the additional circuit inserted in the module as described above.

There are three main factors that can cause the QBER in the QKD system to increase with increasing clock frequency: (1) broadening and patterning of the VCSEL output pulses due to the limited bandwidth of the laser and associated drive electronics; (2) pulse broadening due to dispersion in the fiber; and (3) the timing jitter of the single-photon detectors at the receiver Bob. The most significant contributor to QBER in the system reported here is the detector timing jitter.

Figure 4 shows the improvement in QBER over a range of high clock frequencies from 1GHz to 2GHz. Comparing the standard SPCM module and the enhanced module for a fixed fiber length of 6.55km the QBER drops below 10% between 1 and 2GHz. This is significant, since a QBER of around 10% is regarded as the threshold value below which a quantum key distribution system can be secure from eavesdropping attacks [12]. The impact of the detector modification is evident: at a clock rate of 2GHz, for example, the QBER halves from the prohibitively high figure of ~18% to ~7%, which is below the security threshold. As a result, the estimated net key distribution rate after error correction and privacy amplification [18] improves from zero to the order of 20kbits$^{-1}$ at a transmission distance of 6.55km. At 1GHz the net bit rate and QBER do not differ significantly from the values calculated with the standard SPCM SPAD [9], which are for example, for a transmission distance of 4.2km, a net bit rate of more than 100kbits$^{-1}$ and a QBER of 1.4%.

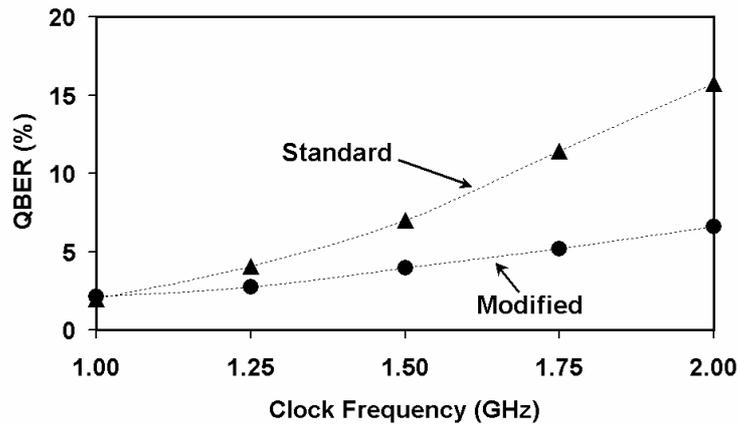

Fig. 4. QBER versus QKD system clock frequency for a fixed fiber length of 6.55 km of standard telecommunications fiber.

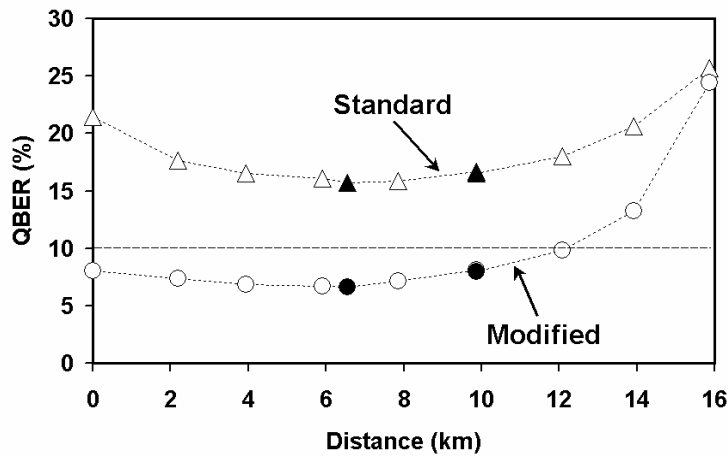

Fig. 5. QBER versus fiber distance at a clock frequency of 2GHz. The points filled in black are taken with the full fiber transmission distance. The white points were measured using optical attenuation to simulate the given distances.

Figure 5 shows QBER versus fiber length at a fixed clock frequency of 2GHz. It is clear that the QBER has dropped to a practical level due to the electronic enhancement in the temporal response of the SPCM module. In addition, the marked increase in QBER that is observed at short distances for the standard detector is greatly reduced with the modified detector due to the reduced temporal broadening at high-count rates.

In summary, these results indicate that use of single-photon detectors with a faster temporal response [19] than the SPCM modules currently used in the QKD system, offer the potential benefits of lower QBER and the consequent advantages of longer distance key distribution and/or higher key exchange rates.

### 3.2. CONCLUSION

The temporal response of a commercially available single-photon counting module has been significantly improved via a relatively low-cost modification consisting of the addition of a single dedicated circuit board. The modified detector has been shown to offer important benefits when applied in a QKD system operating at clock rates in excess of 1GHz. Specifically, the QKD system has been improved in terms of increased workable clock frequency range from 1GHz to

2GHz, with the results at higher frequencies demonstrating the potential for increased transmission distance. For example, for a fiber length of 6.55km and clock rate of 2GHz the QBER was improved from 18% to 7% leading to an increase in potential key distribution rate from zero to 20kbits$^{-1}$. Further improvements in source time response and detector timing resolution will further improve system performance, for example the introduction of faster shallow-junction single-photon avalanche diode detectors [19] and higher bandwidth driving electronics and VCSELs.

## 4. QKD MULTI-USER APPLICATION

An investigation into the use of a 1×32 network [11] – i.e. one transmitter of the initial photon stream and 32 independent potential receivers, with the QKD system described before [9] - was performed. This means that a different cryptographic key can be shared between the transmitter and any of the 32 receivers, but not directly between the each of the 32 receivers. A 1×32 optical splitter/coupler was inserted between the transmitter and receiver, although only four (arbitrarily selected) ports were used for this demonstration (see figure 6). This splitter was designed for a wavelength range of 1520 to 1570nm [9].

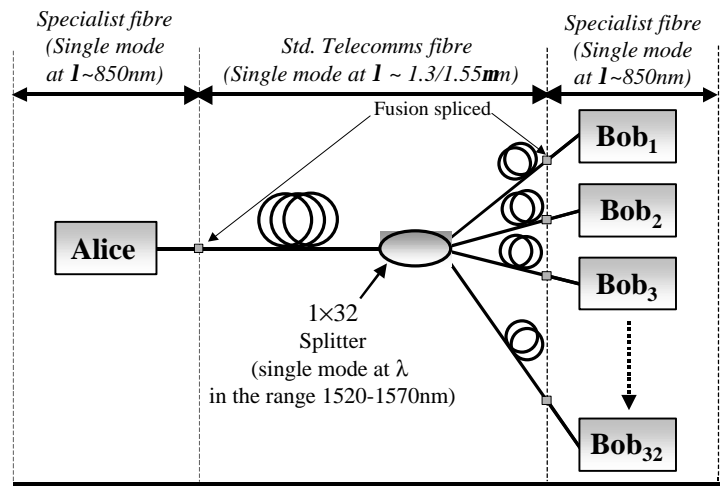

Fig. 6. System schematic. The main transmission span is composed of λ ~ 1.3/1.55μm standard telecommunication fiber components. It is only within the secure send (Alice) and receive (Bob) stations that specialised 850nm fiber is used.

The mean insertion loss of the four ports of the 1×32 splitter at a wavelength of 850nm was measured to be ~18.7dB. This is comparable to the manufacturers' specification of 16.5dB for a wavelength range of 1520 to 1570nm. We must stress that the use of this splitter did incur an increased polarization dependent loss (PDL) when used with 850nm wavelength light. The manufacturers specification quoted a PDL of 0.2dB for a wavelength range of 1520 to 1570nm. However, for a wavelength of 850nm the PDL varied for each port with a maximum measured PDL of 1.1 dB. This meant that the relative angle between the two non-orthogonal linearly polarized states could vary by up to an angle of 5° from the original angle of 45°. This leads to a change in the maximum amount of information that could be gained by Eve [20]. Consequently, for this particular experiment the maximum amount of information that she could gain will increase from the expected 29.3% (corresponding to the expected 45° angle between states) to 36.3% for a 40° angle between states. This change in relative angle will cause small fluctuations in the measured QBER due to the number of detected photons at Bob varying with relative angle. To compensate for this effect in a practical QKD system the current static polarization controllers at Bob should be replaced with automatic polarization controllers. In both PDL and insertion loss measurements, we spliced a specialist 850nm single-mode fiber to the standard telecommunications fiber immediately before and after the splitter. This permitted a more realistic estimate of the PDL, which was originated, at least in part, by the excitation of secondary fiber modes after the splitter.

The 1×32 splitter/coupler was made from a planar lightwave circuit fabricated using an ion exchange process in glass. This fabrication technique exhibits generally a lower PDL compared to fabrication by deposition techniques. This improved PDL performance is due to reduced variation in the density of the inhomogeneous parts in the material during the fabrication process. Also the insertion loss of the 1×32 splitter is less wavelength dependent due to this fabrication process than those using fused biconical taper techniques.

### 4.1. MULTI-USER QKD APPLICATION EXPERIMENTAL RESULTS

Measurements were taken at four of the 32 ports with different lengths of fiber spliced to the arms of the splitter in each case. We estimated the QBER and NBR, after taking into account error correction and privacy amplification [7], for these four distances. These results are shown in Table 1 and Figure 7.

| Fibre Length (km) | QBER (%) | Net Bit Rate/s |
|---|---|---|
| 0.0 | 1.8 | 37516 |
| 2.0 | 2.5 | 20278 |
| 3.8 | 2.8 | 7525 |
| 6.4 | 3.6 | 2640 |

Table 1. QBER and NBR values for the 1×32 system, shown in terms of four point-to-point distances between Alice and each Bob.

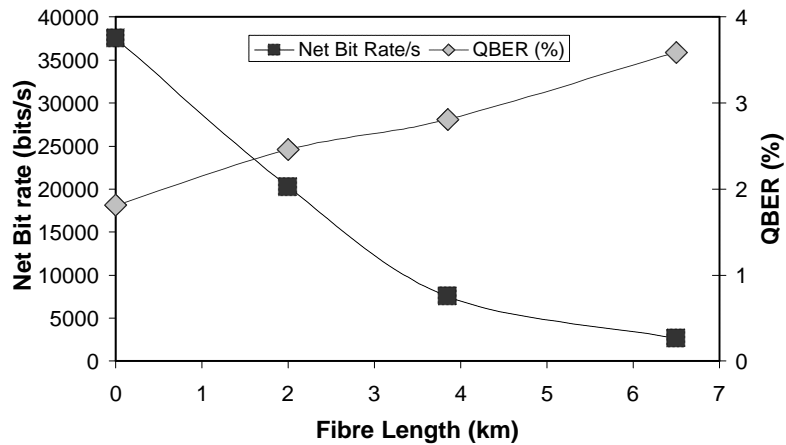

Fig. 7. Graph of QBER and NBR against transmission distance for four of the 32 receivers, each with different fiber length after the output of the 1×32 splitter. Each point of the graph corresponds to a single port.

### 4.2. CONCLUSION

In summary we have demonstrated a quantum key distribution system in a simple multi-user optical network which utilized standard telecommunications optical fiber components as the main transmission medium. The system was clocked at a frequency of 1GHz, and demonstrated a maximum key exchange rate of 37.5kbits$^{-1}$ and a corresponding QBER of 1.8% when used in the 1×32 configuration.

## ACKNOWLEDGMENTS

The authors would like to acknowledge the support of the European Commission SECOQC Integrated Project and the United Kingdom Engineering and Physical Sciences Research Council (project reference GR/N12466). Paul Townsend would like to thank Science Foundation Ireland for support under grant number 03/IN1/1340.

[*] vf2@hw.ac.uk; phone +44-131-451-8060; fax +44-131-451-3136; http://www.phy.hw.ac.uk/resrev/photoncounting